\documentclass[reprint, amsmath,amssymb, aip, apl]{revtex4-1}

\usepackage{graphicx}
\usepackage{dcolumn}
\usepackage{bm}
\usepackage[utf8]{inputenc}
\usepackage[T1]{fontenc}
\usepackage{mathptmx}
\usepackage{etoolbox}
\usepackage[table]{xcolor}
\makeatletter
\def\@email#1#2{%
 \endgroup
 \patchcmd{\titleblock@produce}
 {\frontmatter@RRAPformat}
 {\frontmatter@RRAPformat{\produce@RRAP{*#1\href{mailto:#2}{#2}}}\frontmatter@RRAPformat}
 {}{}
}%
\makeatother
\begin{document}

\preprint{AIP/123-QED}

\title{Large Non-Volatile Frequency Tuning of Spin Hall Nano-Oscillators using Circular Memristive Nano-Gates}

\author{Maha Khademi}
\affiliation{%
 Department of Microtechnology and Nanoscience, Chalmers University of Technology, 412 96, Gothenburg, Sweden.}
\affiliation{NanOsc AB, Kista, Sweden.}
\author{Akash Kumar*}
\affiliation{Applied Spintronics Group, Department of Physics, University of Gothenburg, Gothenburg 412 96, Sweden}
\affiliation{Center for Science and Innovation in Spintronics, Tohoku University, 2-1-1 Katahira, Aoba-ku, Sendai 980-8577 Japan}
\affiliation{Research Institute of Electrical Communication, Tohoku University, 2-1-1 Katahira, Aoba-ku, Sendai 980-8577 Japan}
\author{Mona Rajabali}
\affiliation{NanOsc AB, Kista, Sweden.}
\author{Saroj P. dash}
\affiliation{%
 Department of Microtechnology and Nanoscience, Chalmers University of Technology, 412 96, Gothenburg, Sweden.}%
\author{Johan Åkerman*}
\affiliation{Applied Spintronics Group, Department of Physics, University of Gothenburg, Gothenburg 412 96, Sweden}
\affiliation{Center for Science and Innovation in Spintronics, Tohoku University, 2-1-1 Katahira, Aoba-ku, Sendai 980-8577 Japan}
\affiliation{Research Institute of Electrical Communication, Tohoku University, 2-1-1 Katahira, Aoba-ku, Sendai 980-8577 Japan}
\email{akash.kumar@physics.gu.se and johan.akerman@physics.gu.se}

\date{\today}

\begin{abstract}
Spin Hall nano oscillators (SHNOs) are promising candidates for neuromorphic computing due to their miniaturized dimensions, non-linearity, fast dynamics, and ability to synchronize in long chains and arrays. However, tuning the individual SHNOs in large chains/arrays, which is key to implementing synaptic control, has remained a challenge. Here, we demonstrate circular memristive nano-gates, both precisely aligned and shifted with respect to nano-constriction SHNOs of W/CoFeB/HfO$_{x}$, with increased quality of the device tunability. Gating at the exact center of the nano-constriction region is found to cause irreversible degradation to the oxide layer, resulting in a permanent frequency shift of the auto-oscillating modes. As a remedy, gates shifted outside of the immediate nano-constriction region can tune the frequency dramatically ($>$200 MHz) without causing any permanent change to the constriction region. Circular memristive nano-gates can, therefore, be used in SHNO chains/arrays to manipulate the synchronization states precisely over large networks of oscillators. 

\end{abstract}

\maketitle


\section{\label{sec:level1}Introduction}

Spin Hall nano oscillators (SHNOs), operating based on the principle of the spin Hall effect~\cite{Dyakonov1971,Hirsch1999,Sinova2015Rev}, are nano-meter-sized microwave signal-generating devices composed of a ferromagnetic layer adjacent to a heavy metal layer~\cite{VEDemidov2012,Liu2012PRL,duan2014nanowire,demidov2014nanoconstriction}. Their large non-linearity, fast dynamics, and ability to synchronize in chains and arrays~\cite{awad2016natphys,zahedinejad2020two,kumar2023robust}, make SHNOs promising candidates for emerging applications
including microwave signal processing~\cite{muduli2010nonlinear}, ultrafast spectral analysis~\cite{litvinenko2020ultrafast,litvinenko2022ultrafast}, and unconventional computing~\cite{zahedinejad2020two,Romera2018nt,houshang2022prappl,Torrejon,Manipatruni,Yogendra}. Among SHNOs, the nano-constriction (NC) geometry attracts the most attention due to its easy fabrication and direct optical access~\cite{demidov2014nanoconstriction}. For oscillator based computing, it is necessary to tune the properties of individual SHNOs in long chains or large arrays. Several attempts have been made in this direction, such as opto-thermal ~\cite{Muralidhar2022optothermal} and voltage control~\cite{fulara2020giant,Choi2022,kumar2022fabrication}. Recently, Zahedinejad \textit{et al.} demonstrated memristive control of mutual synchronization in SHNO chains (using metal oxide memristors on top of SHNOs) and the results showed overall tunability of 60 MHz~\cite{Zahedinejad2022natmat}. The memristors provide non-volatile control of the SHNOs that may be used for neuromorphic computing, along with on-chip memory for training purposes. However, the use of a wide bar-shaped gate over the NC and its edges required a relatively thick oxide layer (greater than 19 nm) to prevent undesired sidewall shunting, restricting its usefulness and scalability for large arrays while also reducing the frequency tunability. 
\begin{figure}[t!]
\centering
\includegraphics[width=250pt]{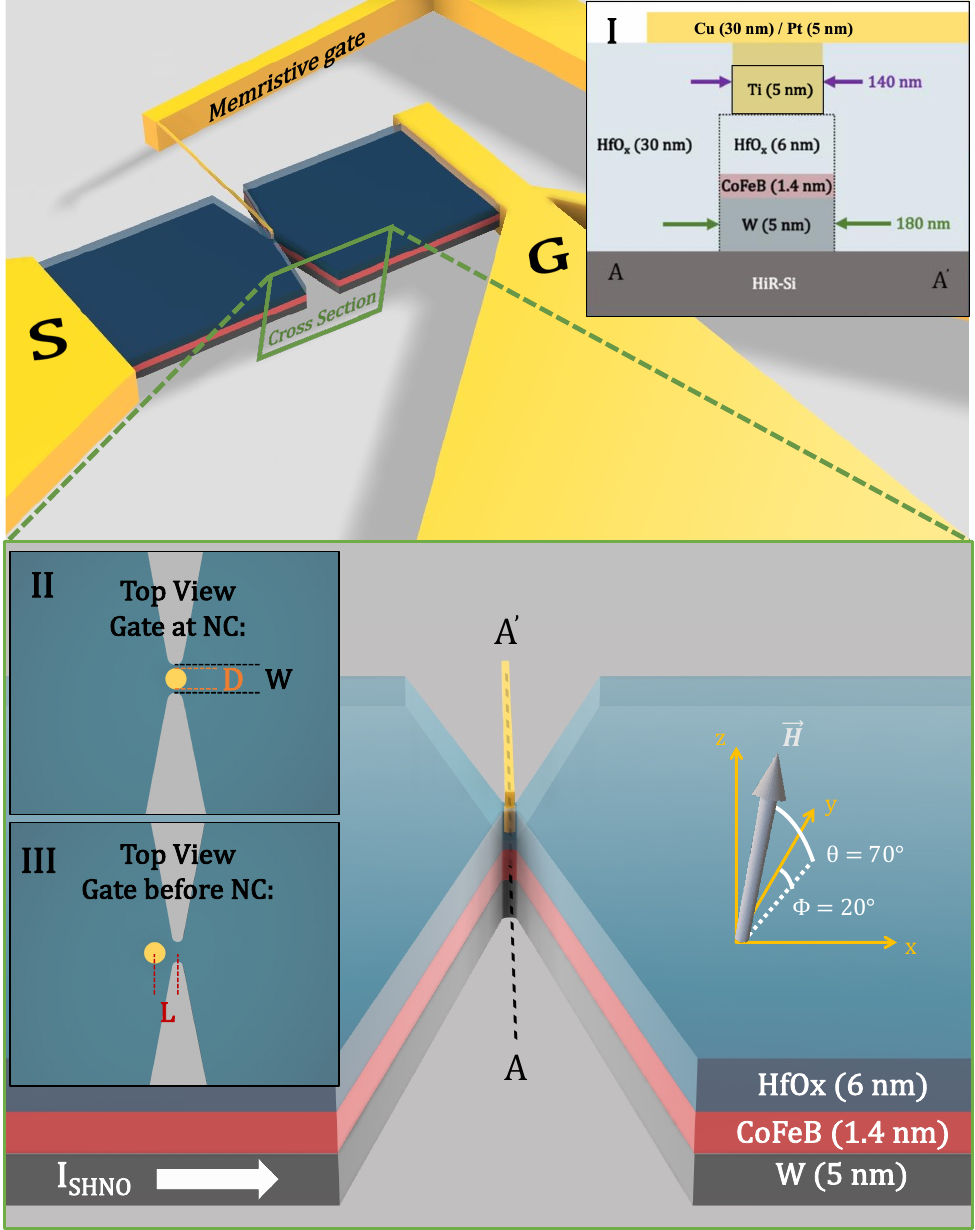}
   \caption{\label{fig:Fig1} Schematic representation of the gated SHNO with the memristor, signal (S), and ground (G) contact pads. A cross-sectional view of the SHNO at the external magnetic field is provided. Inset I: Material stacks in AA' cross-section, showing the order and the thickness of the material. All of the stacks are encapsulated with 30 nm HfO$_{x}$. Inset II: Top view of a gate at the nano-constriction region with D=140~nm and W=180~nm. Inset III: Top view of a gate before the nano-constriction with D=200~nm, W=180~nm and L=200~nm.} 
\end{figure}

Recently, we demonstrated the successful fabrication of precisely aligned circular nano-gates for voltage control of W/CoFeB/MgO based SHNOs~\cite{kumar2022fabrication}. The refined fabrication process not only enables precise positioning but also eliminates the need for thick oxide layers between the gate and the ferromagnetic layers. In this work, we fabricate and compare

\noindent circular memristive nano-gates positioned on top of, and just next to, the NC SHNOs of W/CoFeB/HfO$_{x}$. By substituting MgO with HfO$_{x}$ (while preserving the magnetic properties), it becomes possible to further reduce the thickness of the oxide layer and mitigate the uncertainties in switching that were previously associated with the use of mixed oxide 
layers.~\cite{fulara2020giant,Zahedinejad2022natmat,kumar2022fabrication,Choi2022}. To achieve memristive functionality, we employ 5 nm Ti thin films instead of the previously used Pt thin films~\cite{kumar2022fabrication}. We observe that gates on top of the NC cause permanent degradation and result in weak frequency tunability, whereas gates next to the NC lead to a strong frequency tunability without changing the magnetodynamic properties of the SHNOs~\cite{Zahedinejad2022natmat}. 

\vspace{-6.2mm}

\section{\label{sec:level2}Experiment Details}

Nano-constriction SHNOs were fabricated using magnetron sputtered W(5~nm)/CoFeB(1.4~nm)/HfO$_{x}$(6~nm) thin film stacks on an intrinsic high resistivity Si substrate ($\rho_{Si}>$10~k$\Omega$cm). The thin film stack was annealed in ultra-high vacuum conditions at 300 $^\circ$C for one hour to induce PMA. Compared to previous works~\cite{fulara2020giant,Zahedinejad2022natmat,kumar2022fabrication,Choi2022}, the thin film stacks of W/CoFeB/HfO$_{x}$ were optimized without MgO while maintaining low Gilbert damping ($\approx$0.025) and large PMA field ($>$1 T)~\cite{Gaur_spin_pumping}. 
This not only minimizes the overall oxide thickness but also allows the direct fabrication of gates on a single oxide layer, which, in turn, eliminates uncertainties and interfacial effects arising from the use of different oxide materials in the dielectric layers.
The SHNO mesa, featuring nano-constriction widths of 180 nm, was produced by combining e-beam lithography (EBL, Raith EBPG 5200) and Ar-ion etching techniques. Chip markers were included for precise alignment. Following the fabrication of the mesa, a 30~nm HfO$_{x}$ thin film was deposited using RF magnetron sputtering to encapsulate the gate electrode. An EBL process, followed by Ar-ion etching, was used to create a through-hole as the gate. We utilize a 5 nm Ti (along with 30 nm Cu and 5 nm Pt) thin film as the top electrode material which is well known for its memristive properties along with HfO$_{x}$ resulting in conduction filament formation as the resistive switching mechanism.~\cite{Lubben,Wedig2016,zhang2021evolution} The high resistance state (HRS) and low resistance state (LRS) are characterized by resistance levels in the orders of a few M$\Omega$ and several k$\Omega$, respectively.
Here, thick Cu/Pt layers were utilized to minimize contact resistance.
The gates were precisely aligned to the center of nano-constriction with the help of precise chip alignment in e-beam lithography. As a result, a 140~nm circular gate was fabricated on top of the 180~nm nano-constriction as shown in Fig.~\ref{fig:Fig1}.
Thanks to the large spin Hall angle of W thin films ($>-$0.44), the threshold current density for these devices are found to be about 0.55 mA~\cite{mazraati2016low,behera2022energy}.
A cross-sectional schematic image of the material stacks is shown in Fig.~\ref{fig:Fig1} inset I. Top views for two different geometries discussed in this paper are shown in insets II and III. 
The detailed fabrication process can be found here.~\cite{kumar2022fabrication}

\section{Results and Discussion}

\begin{figure}[t!]
\centering
\includegraphics[width=250pt]{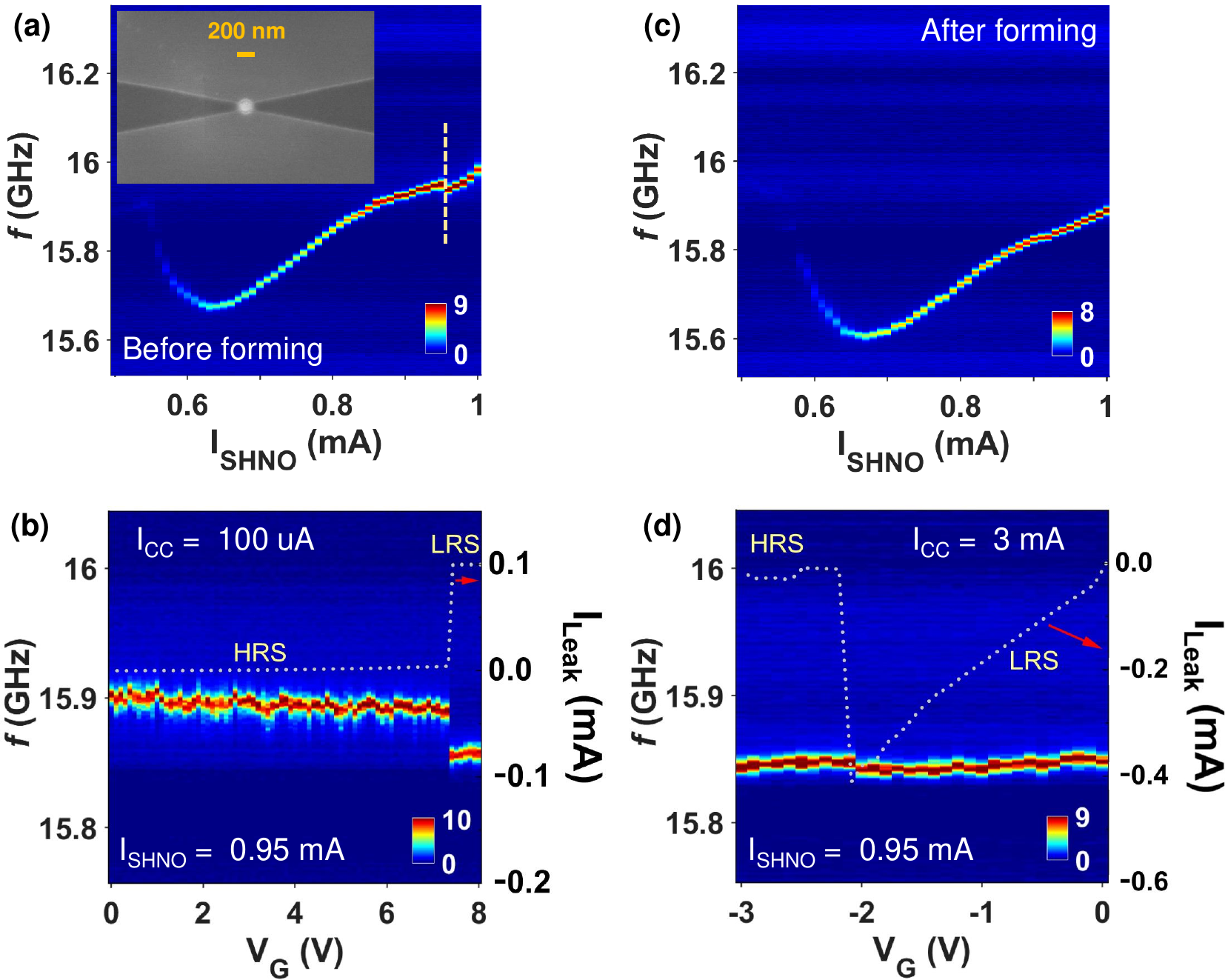}
   \caption{\label{fig:Fig2} (a) PSD \emph{vs.}~$\mathit{I_{SHNO}}$ of a 180~nm NC with a 140~nm circular gate on top, before forming the memristive gate; a 0.8~T magnetic field was applied with $70^\circ$ out-of-plane and $20^\circ$ in-plane angle. The dashed line indicates the current at which gating is done. Inset: SEM image of the gated SHNO. The frequency and leakage current \emph{vs.}~gate voltage for (b) 0 to 8~V (forward sweep) and (d) 0~V to -3~V (reverse sweep). (c) PSD of the auto-oscillation after forming the memristive gate.}
\end{figure} 

\textit{Gate on top of the nano-constriction:} Figure~\ref{fig:Fig2}(a) shows the current dependence of the auto-oscillation (AO) power spectral density (PSD) observed for SHNOs with a nano-gate at the nano-constriction (before forming the memristor), the inset shows the scanning electron micrograph (SEM) image for the same. The gate voltage is swept with a compliance current (I$_{\rm CC}$) of 100 $\mu $A [Fig.~\ref{fig:Fig2}(b)]. 
The compliance current is a key parameter for memristors as it controls the size of the conducting path in the insulating layers and can result in multi-level resistance states, which are highly useful for neuromorphic computing~\cite{he2017customized}. This also safeguards the device from breakdown and does not allow any additional voltage/current in the system.
Starting from the HRS (10.17 M$\Omega$) and increasing the gate voltage, the memristor switches (forms) into its LRS (1.37 $k\Omega$) at 7.5~V (forming voltage) (Fig.). The forming of the memristor results in a 30 MHz irreversible downward shift [Fig.~\ref{fig:Fig2}(b)]. This might be due to permanent degradation in the material in the nano-constriction region underneath the oxide layer. We believe this permanent degradation changes the interfacial perpendicular magnetic anisotropy (iPMA) under the gate, and as a result, shifts the AO signal 30 MHz down [Fig.~\ref{fig:Fig2}(c)].
It has been previously reported~\cite{diezinterface} that adding a dusting layer between the CoFeB and the oxide layer would directly change the electronic states at the interface and affect the anisotropy of the system. We believe devices with a gate at the NC suffer from a similar ion migration due to memristive filament formation which eventually changes the frequency of operation. 
Repeating the same measurement on several devices reveals the random nature of degradation as we observe a device-to-device varying frequency shift (20-60 MHz). It is noteworthy that such permanent degradations were not reported previously due to the large oxide thicknesses and surface contact between the gate and the nano-constriction region~\cite{Zahedinejad2022natmat}.
Moreover, it is observed that in three-terminal devices such as gated-SHNOs when the gate is at the middle of the NC, the leakage current (memristive current) from gates only passes through half of the SHNOs~\cite{Albertsson_3terminalSHNO}.

\begin{figure}[t!]
\centering 
\includegraphics[width=250pt]{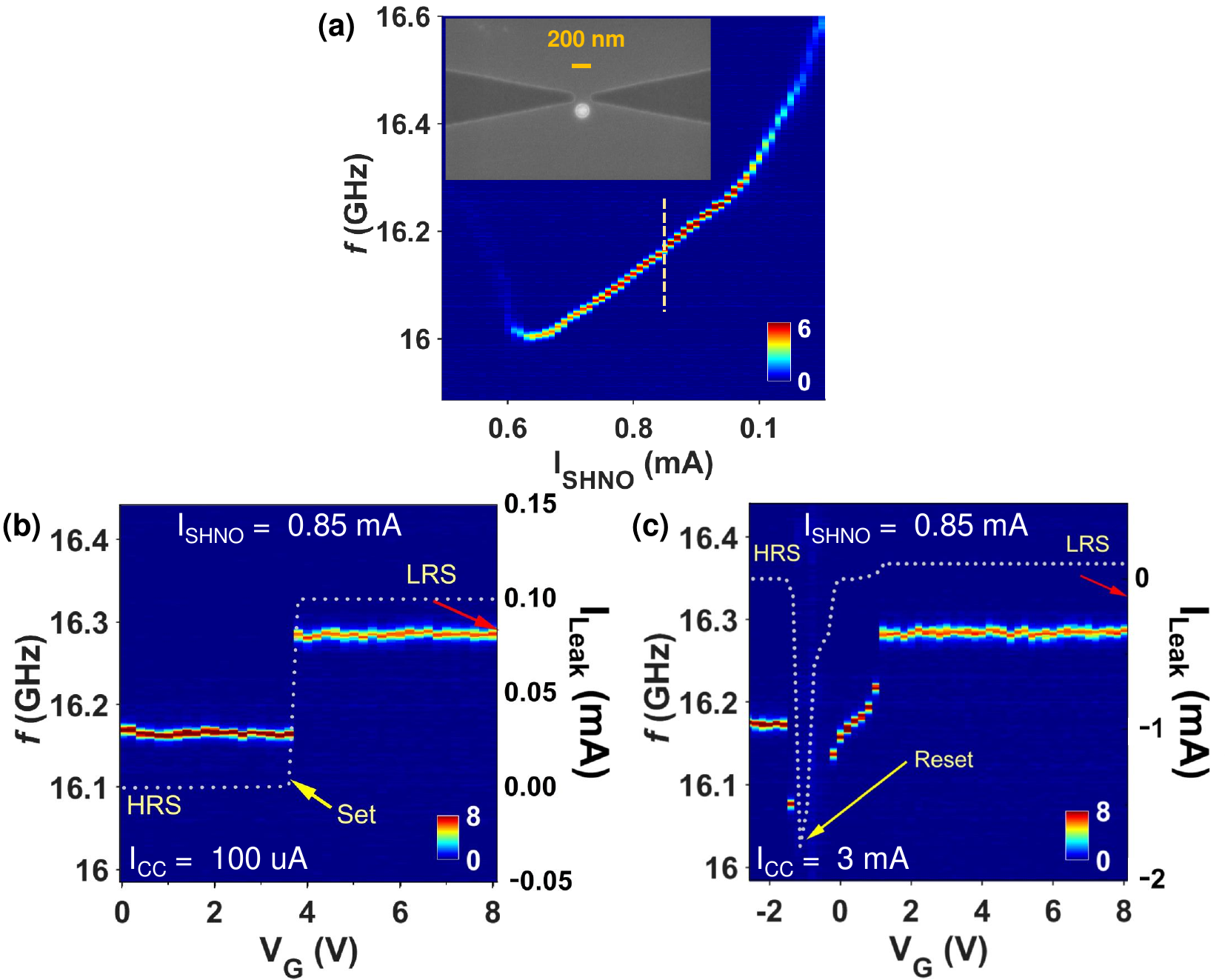}
\vspace{-0.5cm} 
   \caption{\label{fig:Fig3} The PSD plot \emph{vs.}~$\mathit{I_{SHNO}}$ at the same field configurations as in Fig.~\ref{fig:Fig2} for a 180~nm nano-constriction with a 200~nm circular gate separated by 200~nm. The dashed line indicates the current at which the gating is done. Inset: SEM image of a similar gated SHNO. The frequency and leakage current \emph{vs.}~gate voltage from (b) zero to 8~V (forward sweep) and (c) 8~V to -3~V (reverse sweep).}  
\end{figure}

In the reverse sweep [Fig.~\ref{fig:Fig2}(d)], the memristive element switches back (resets) to its HRS at -2.2~V, and the leakage current drops to zero. However, we do not observe any change in the AO frequency. This reveals that the memristor is not functional anymore. As mentioned earlier, it might be due to local permanent changes under the gate. Hence, we can conclude that placing the gate on top of the NC causes degradation and does not produce the desired frequency tunability.

\textit{Gate shifted away from the nano-constriction:} To mitigate the detrimental effect of degradation, we fabricate and study 200~nm circular gates placed 200 nm away from the NC, along the current direction. The SEM image of the fabricated device, together with its AO PSD plot, is shown in Fig.~\ref{fig:Fig3}(a). During forward voltage sweep, the AO frequency shows a frequency shift of almost 130~MHz at 4~V (forming voltage) (equivalent to 100 $\mu $A current though the memristive gate), and the leakage current rises as the gate sets into its LRS [Fig.~\ref{fig:Fig3}(b)]. The reason for this lies behind the current density distribution as the memristor in its LRS state adds current to the NC. Thus, the AO mode at the NC experiences a higher current density, resulting in a large frequency tunability. In the reverse voltage sweep, the gate resets to the HRS at -1.6~V, the leakage current drops to zero (with compliance current of 3~mA), and the oscillator ends up in
the same frequency as before [Fig.~\ref{fig:Fig3}(c)]. This indicates that any degradation happens outside of the magneto-dynamically active region and, hence, their impact on the oscillatory behavior remains minimal. It is worth noting that there are no restrictions on the maximum distance for positioning the gate ~\cite{victor_voltagecontrol_sim, dvornik2018origin}. A frequency tunability of more than 200 MHz can be obtained in the reverse sweep between +1 V to -1.6 V. The observed frequency shift is significantly larger than in our previous work with broad gates~\cite{Zahedinejad2022natmat}. 

The fabrication of gates before the NC also provides an opportunity to fabricate multiple memristor gates (or memristor arrays) for oscillators, which can provide tunable synaptic weights for each NC~\cite{moradi2019spin}. Apart from neuromorphic computing the large frequency tunability of these devices can be used for energy efficient electronics such as tunable GHz frequency antennas~\cite{chen2014highly}, tunable energy harvesting devices using large chains of SHNOs with memristive gating~\cite{sharma2021electrically}, signal processing using ultrafast spectrum analyzer~\cite{litvinenko2022ultrafast} and highly sensitive magnetic field sensors.~\cite{albertsson2020magnetic}.

\section{\label{sec:level4}Conclusion}

In summary, we demonstrate the implementation of circular memristive nano-gates in two different configurations: on top of and next to nano-constriction Spin-Hall Nano-Oscillators (SHNOs). Our findings reveal that placing the gate on top of the nano-constriction leads to changes in the material properties and adversely affects the frequency tunability. Conversely, when the gates are positioned 200 nm away from the nano-constriction, it prevents any such damage to the nano-constriction. Furthermore, due to higher current densities from memristive current in the area surrounding the nano-constriction, this configuration enables a substantial frequency tunability exceeding 200 MHz. These characteristics hold significant potential for controlling and manipulating individual SHNOs within chains and arrays, facilitating precise regulation of the mutual synchronization among them. 

\begin{acknowledgments}
This project has received funding from the European Union’s Horizon 2020 
research and innovation programme under the Marie Skłodowska-Curie grant 
agreement No 955671 “SPEAR”, Grant agreement No. 899559 “SpinAge”, and Advanced Grant No. 835068 “TOPSPIN”. This work was also partially supported by the Swedish Research Council (VR Grant No. 2016-05980) and the Knut and Alice Wallenberg Foundation.
\end{acknowledgments}

\bibliography{Main}

\end{document}